\documentclass[preprint,showpacs,amsmath,amssymb,tightenlines]{revtex4}
\usepackage{graphicx}
\usepackage{color}

\newcommand{\dd}{\mathrm{d}}      
\newcommand{\pd}[2]{\frac{\partial #1}{\partial #2}} 
\newcommand{\td}[2]{\frac{\dd #1}{\dd #2}} 
\newcommand{\mean}[1]{\left\langle #1 \right\rangle} 
\newcommand{\IInt}[3]{\int_{#2}^{#3}\dd #1\;} 
\newcommand{\Int}[1]{\int\dd #1\;} 

\newcommand{\ps}{p^\text{s}}
\newcommand{\eps}{\varepsilon}
\newcommand{\ex}{_\mathrm{ex}}
\DeclareMathOperator{\tr}{tr}
\DeclareMathOperator*{\reaction}{\rightleftharpoons}
\newcommand{\x}{\mathbf{x}}
\newcommand{\nt}{\boldsymbol\zeta}
\newcommand{\J}{\mathbf{J}}
\newcommand{\M}{\mathbf{M}}
\newcommand{\NN}{\mathcal{N}}

\def\la{\lambda(\tau)}
\def\l{\lambda}

\def\dv{{\partial V\over\partial x}}

\def\la{\lambda(\tau)}
\def\l{\lambda}

\def\dsm{\dot s_{\rm m}(\tau)}
\def\m{_{\rm m}}
\def\t{_{\rm tot}}

\def\pnt{p_{n(\tau)}(\tau)}

\def\wpm{w_{n_j^+n_j^-}}
\def\wmp{w_{n_j^-n_j^+}}
\def\wpmb{w_{\n_j^+\n_j^-}}
\def\wmpb{w_{\n_j^-\n_j^+}}
\def\njp{{\n_j}^+}
\def\njm{{\n_j}^-}

\def\c{{\cal C}}

\def\tot{_{\rm tot}}
\def\m{_{\rm m}}

\def\dv{{\partial V\over\partial x}}

\def\wc{w_{\rm chem}}

\def\r{r_{\alpha}^{nm}}
\def\s{s_{\alpha}^{nm}}
\def\d{(\r-\s)}
\def\c{c_\alpha}
\def\ceq{\c^{\rm eq}}
\def\ceqa{(\c^{\rm eq})^{\r-\s}}
\def\A{A_\alpha}
\def\wnm{w_{nm}}
\def\wmn{w_{mn}}
\def\wnmo{w_{nm}^0}
\def\wmno{w_{mn}^0}
\def\Wnm{W_{nm}(\n)}
\def\Wmn{W_{mn}(\n')}

\def\xal{\x^\alpha}
\def\mij{\mu_{ij}}

\def\dv{\partial V \over \partial}

\def\E{E_\alpha}

\def\meq{\mu_\alpha^{\rm eq}}
\def\m{\mu_\alpha}

\def\sa{\sum_\alpha}

\def\n{{\bf n}}

\begin{document}

\title{Entropy production for mechanically or chemically driven biomolecules}
\author{Tim Schmiedl, Thomas Speck, and Udo Seifert}

\affiliation{{II.} Institut f\"ur Theoretische Physik, Universit\"at Stuttgart,
  70550 Stuttgart, Germany}

\begin{abstract}
  Entropy production along a single stochastic trajectory of a biomolecule is
  discussed for two different sources of non-equilibrium. For a molecule
  manipulated mechanically by an AFM or an optical tweezer, entropy production
  (or annihilation) occurs in the molecular conformation proper or in the
  surrounding medium. Within a Langevin dynamics, a unique identification of
  these two contributions is possible. The total entropy change obeys an
  integral fluctuation theorem and a class of further exact relations, which
  we prove for arbitrarily coupled slow degrees of freedom including
  hydrodynamic interactions. These theoretical results can therefore also be
  applied to driven colloidal systems. For transitions between different
  internal conformations of a biomolecule involving unbalanced chemical
  reactions, we provide a thermodynamically consistent formulation and
  identify again the two sources of entropy production, which obey similar
  exact relations. We clarify the particular role degenerate states have in
  such a description.
\end{abstract}

\maketitle


\section {Introduction}

Biological systems are generically out of equilibrium. Still, for most
processes in cell biology taking place on the level of a single (or few)
molecules, the intracellular aqueous solution provides an environment with
constant temperature. The genuine source of non-equilibrium are not
temperature gradients but rather mechanical or chemical stimuli provided by
external forces or imbalanced chemical reactions. Such a characterization
motivates the quest for a thermodynamical understanding of mechanically or
chemically driven non-equilibrium processes taking into account their
necessarily stochastic character on the level of few molecules~\cite{bust05}.
Crucial for such a program are consistent formulations of the first and the
second law under these conditions.

For mechanically driven processes, the controlled unfolding of proteins, RNA,
and DNA typically described by Langevin equations can serve as a paradigm (for
a review, see Ref.~\cite{rito02}). For the overdamped motion of a single
colloidal degree of freedom, Sekimoto has shown how to relate work, internal
energy and exchanged heat with the terms occurring in the Langevin equation,
thus providing a formulation of the first law on the level of a single
trajectory~\cite{seki98}. The extension of this interpretation to a
biomolecule with several overdamped spatial degrees of freedom subject to both
a potential of mean force and some additional mechanical force applied via an
AFM or optical tweezers is, in principle, straightforward and will be given
below. As a refinement of the second law, the Jarzynski relation expresses the
free energy difference of an initial (folded) and a final (unfolded) state by
an exponential average of the non-equilibrium work spent in such a
transition~\cite{jarz97,jarz97a,croo99,croo00}. This relation has found
wide-spread attention both in experimental and theoretical studies of
unzipping and unfolding
transitions~\cite{humm01,liph02,brau04,park04,coll05,spec05}. It has also
inspired theoretical studies on the probability distribution of the work spent
in such processes~\cite{spec04,impa05,impa05a}. Even though the Jarzynski
relation does not explicitly require a definition of entropy on the level of a
single trajectory, one obtains a second-law like inequality for the average
work as a mathematical consequence. The concept of an entropy of a single
trajectory is fruitful since it allows to derive equalities different from but
related to the Jarzynski relation for the total entropy change
directly~\cite{seif05a}.

For chemically driven processes, an equally comprehensive understanding and
formulation is not yet available. Based on classical work on network
thermodynamics~\cite{oste71,schn76,hill,nicolis}, ensemble properties like
mean heat dissipation or entropy production rate have been identified and
investigated (see~\cite{gasp04,andr04,qian05,qian05a} and references therein)
with only a few attempts to provide a thermodynamic interpretation of the
single reaction events~\cite{shib00,seif05}. Taking the Langevin equation for
mechanically driven processes as a guideline, however, it should be possible
to formulate for single biochemical reaction events a first-law like energy
conservation statement. Likewise, for a proper formulation and refinement of
the second law, one should develop a notion of entropy along such a single
stochastic history of reaction events. Only after averaging one will then
recover previous ensemble formulations. The motivation for such a
trajectory-based approach also derives from the exciting experimental
possibilities to study conformational changes of single enzymes using
fluorescence spectroscopy as reviewed in Refs.~\cite{schw01,xie02}. Finally,
molecular motors comprise a class of systems where biochemical reactions lead
to discrete mechanical steps for which such a thermodynamic modeling should
become appropriate as
well~\cite{qian97,juel97,fish99,lipo00,qian01,reim02a,maes03a,bake04}.

This paper presents a coherent theoretical framework for describing both
mechanically or chemically driven transitions between different
configurational internal states of a biomolecule in a way that is
thermodynamically consistent on the level of a single trajectory. In
particular, concerning entropy production, we exploit the general framework
introduced in Ref.~\cite{seif05a} for such isothermal non-equilibrium
processes. In Section~\ref{sec:mech}, we consider the mechanically driven
dynamics of a biomolecule involving several (slow) degrees of freedom. We
provide a first law-like interpretation of the Langevin equation for its
coupled overdamped degrees of freedom and derive exact relations on entropy
production along such a driven trajectory, thereby extending our previous work
both to many degrees of freedom and to (long-range) hydrodynamic interactions
among them. Such interactions will become particularly relevant for colloidal
systems (to which the same formalism is applicable) if extant studies of their
non-equilibrium
thermodynamics~\cite{hata01,wang02,carb04,zon03a,zon03,trep04,blic05a} are
pushed beyond the one particle level. In Section~\ref{sec:chem}, we first
consider transitions between different internal states of a protein or enzyme
caused by biochemical reactions involving unbalanced chemical species which
are the source of non-equilibrium in this case. We then apply the general
notion of entropy production introduced in Ref.~\cite{seif05a} to such
transitions and derive exact relations for the total entropy production.
Finally, we discuss the modifications arising from a possible degeneracy of
the states occurring in such a description. In Section~\ref{sec:persp}, we
discuss a few perspectives of our approach. The Appendix contains the
path-integral based proof of a general integral fluctuation theorem for
(hydro)dynamically coupled degrees of freedom in a time-dependent potential.


\section{Mechanically driven case}
\label{sec:mech}

\begin{figure}[t]
  \centering
  \includegraphics[width=7cm]{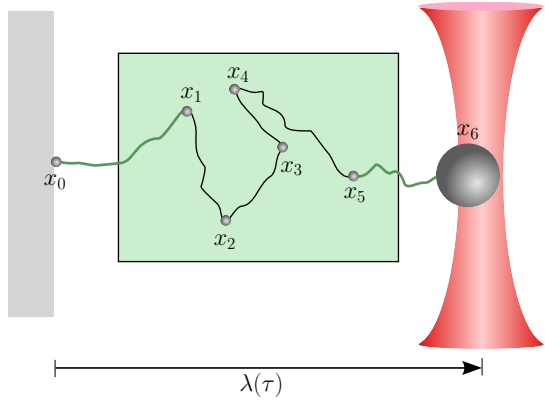}
  \caption{Biomolecule with (slow) degrees of freedom $\x=(x_1,\dots,x_5)$
    attached via polymeric linkers to a substrate ($x_0$, left end) and a bead
    ($x_6$, right end) controlled externally by laser tweezers at position
    $\lambda(\tau)$. The bare potential $V_0(\x)$ involves the internal
    degrees of freedom. The external potential can be modeled as
    $V\ex(\x,\l)=V_1(x_1-x_0)+V_1(x_6-x_5)+(k/2)[x_6-\lambda(t)]^2$, where
    $V_1(y)$ is the potential for a (semi-flexible) linker with extension $y$
    and $k$ is the strength of the optical trap.}
  \label{fig:mol}
\end{figure}

We describe the biomolecule by a set $\x\equiv(x_1,\dots,x_d)$ of internal
coordinates, which should comprise the relevant $d$ slow degrees of freedom.
In equilibrium, this molecule feels a potential (of mean force) $V_0(\x)$.
Optical tweezers or a cantilever attached via a linker give rise to an
additional potential $V\ex(\x,\l)$. The external control parameter $\la$
describes the time-dependent motion of the tweezer focus or the base of the
cantilever, see Fig.~\ref{fig:mol}. As equation of motion, we choose a
Langevin description
\begin{equation}
  \dot x_i = -\mij\pd{V}{x_j} + \zeta_i,
  \label{eq:lang}
\end{equation}
where summation over repeated indices is understood throughout the paper.
Here $V(\x,\l)\equiv V_0(\x)+V\ex(\x,\l)$ is the sum of both potentials. We
allow for a non-diagonal mobility $\mij(\x)$ which can include hydrodynamic
interactions, e.g., through an Oseen tensor~\cite{doiedwards}. The stochastic
increments $\zeta_i$ are modeled as Gaussian white noise with
\begin{equation}
  \langle \zeta_i(\tau) \zeta_j(\tau')\rangle \equiv
  2\mij(\x)\delta(\tau-\tau').
  \label{eq:corr}
\end{equation}
Throughout the article, we measure energies in units of $k_\mathrm{B}T$, which
is set to 1. Likewise, we use a dimensionless entropy, i.e., we set the
Boltzmann constant $k_\mathrm{B}$ to 1 as well. Under equilibrium conditions
for constant $\l$, the type of correlations (\ref{eq:corr}) guarantees that
the Boltzmann distribution $p(\x,\l)\sim \exp[-V(\x,\l)]$ is stationary. It is
an essential assumption for the theory we will be discussing that these
correlations persist despite the fact that for a time-dependent protocol $\la$
we are no longer in equilibrium.

The Langevin dynamics can be cast in the form of the first law, i.e., energy
conservation along a stochastic trajectory~\cite{seki98}. Manipulating the
system by changing the external control parameter $\l$ gives rise to an
increment in applied work
\begin{equation}
  \dd w \equiv \pd{V}{\lambda}\dd\lambda.
\end{equation}
This work will either change the internal energy
\begin{equation}
  \dd V = \pd{V}{x_i}\dd x_i + \pd{V}{\lambda}\dd\l
\end{equation}
or is dissipated as heat
\begin{equation}
  \dd q = \dd w - \dd V = -\pd{V}{x_i}\dd x_i
\end{equation}
into the thermal environment. Since the heat bath has constant temperature, we
can identify this exchanged heat with a change in entropy of the medium as
\begin{equation}
  \label{eq:sm:mech}
  \dsm = \td{q}{\tau} = -{\dv x_i}\dot x_i.
\end{equation}
This quite natural definition of the entropy change of the medium along each
trajectory raises the question whether there is a corresponding entropy change
of the biomolecule itself.

Following the route outlined in Ref.~\cite{seif05a}, we now show that such an
entropy of the ``system'' can consistently be defined along each stochastic
trajectory $\x(\tau)$ as
\begin{equation}
  s(\tau)\equiv -\ln p(\x(\tau),\tau),
\end{equation}
where $ p(\x,\tau)$ is the solution of the Fokker-Planck equation for
the probability distribution
\begin{equation}
  \partial_t p(\x,\tau) =-\partial _ij_i(\x,\tau) = \pd{}{x_i}\mij\left[
    \pd{V}{x_j} + \pd{}{x_j} \right]p(\x,\tau).
\end{equation}
Upon averaging with $p(\x,\tau)$, this stochastic entropy becomes the
non-equilibrium Gibbs or Shannon entropy
\begin{equation}
  S(\tau)\equiv \langle s(\tau)\rangle = -\Int{^dx} p(\x,\tau) \ln p(\x,\tau).
\end{equation}
The advantage of defining such a system entropy is that one can proof quite
general theorems involving the total entropy change
\begin{equation}
  \Delta s\t \equiv s(t)-s(0) + \IInt{\tau}{0}{t}\dsm
\end{equation}
along a stochastic trajectory $\x(\tau)$ of length $t$. As shown in
Appendix~\ref{sec:ft}, this total entropy change obeys the integral
fluctuation theorem
\begin{equation}
  \label{eq:ft}
  \mean{\exp[-\Delta s\t]} = 1,
\end{equation}
which implies immediately the second law in the form
\begin{equation}
  \mean{\Delta s\t} \geq 0.
\end{equation}
The brackets $\mean{\cdots}$ denote the average over infinitely many
realizations of the process. Moreover, for any function of the final
coordinates $f(\x_t)$ one even has the relation
\begin{equation}
  \label{eq:ft:cro}
  \mean{f(\x_t)\exp[-\Delta s\t]} = \mean{f(\x_t)}.
\end{equation}
The relations~\eqref{eq:ft} and~\eqref{eq:ft:cro} are quite universal since
they hold for the non-equilibrium average $\mean{\cdots}$ with any initial
distribution $p(\x,0)$, for any trajectory length $t$, and for any driving
protocol $\l(\tau)$.

These relations should be distinguished from both the Jarzynski
relation~\cite{jarz97,jarz97a}
\begin{equation}
  \label{eq:ft:jarz}
  \mean{\exp[-W_{\rm d}]} = 1
\end{equation}
and the relation~\cite{croo00}
\begin{equation}
  \label{eq:ft:crooks}
  \mean{f(\x_t)\exp[-W_{\rm d}]} = \mean{f(\x_t)}_{{\rm eq},\l(t)},
\end{equation}
where $W_{\rm d}\equiv W-\Delta F=W-[F(\l(t))-F(\l(0))]$ is the dissipated
work involved in the non-equilibrium transition between the initial
equilibrium state at $\l(0)$ with free energy $F(\l(0))$ and the final state
at $\l(t)$ with free energy $F(\l(t))$. In particular, in
relation~\eqref{eq:ft:crooks} the average on the right hand side corresponds
to an equilibrium average at the final value of the control parameter, whereas
in~\eqref{eq:ft:cro} it is the average involving the actual probability
distribution $p(\x,t)$. It is crucial to note that for Eqs.~\eqref{eq:ft:jarz}
and~\eqref{eq:ft:crooks} the initial distribution has to be the thermal
equilibrium distribution for $\l(0)$ whereas in Eqs.~\eqref{eq:ft}
and~\eqref{eq:ft:cro} it is arbitrary.

Even though the motivation of this presentation is on biomolecules, it should
be clear that the mechanically driven case discussed here applies exactly to
colloidal particles coupled through direct or hydrodynamically induced
interactions and driven by time-dependent laser traps. For such systems, these
theorems show that fluctuation theorems (as well as the Jarzynski relation)
persist in the presence of hydrodynamic interactions.


\section{Chemically driven case}
\label{sec:chem}

\subsection{Enzyme or protein with internal states}

\begin{figure}
  \includegraphics[width=6cm]{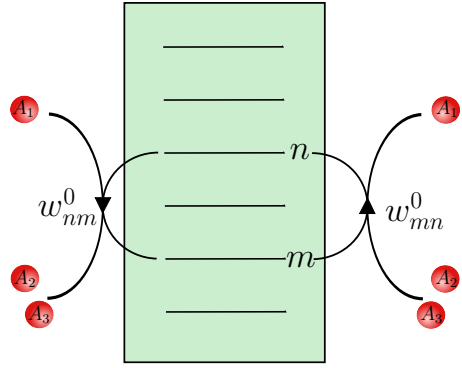}
  \caption{Protein or enzyme with internal states. A forward transition (left)
    from $n$ to $m$ involves the chemical reaction $A_1 + n \rightarrow m +
    A_2 + A_3$ and similarly for the backward reaction (right). The rates
    $w^0_{nm}$ and $w^0_{mn}$ are the (not concentration-dependent) bare
    rates.}
  \label{figure2}
\end{figure}

As a model for a biomolecule driven by chemical forces, we consider a protein
with $M$ internal states $\{1,2, ...,M\}$. Each state $n$ has internal energy
$E_n$. Transitions between these states involve some other molecules $\A$,
where $\alpha=1,\ldots,N_A$ labels the different chemical species. We assume
that the chemical potentials, i.e., the concentrations $\c$ of these molecules
are controlled or clamped externally. A transition from state $n$ to state $m$
implies the reaction
\begin{equation}
  \sa\r\A + n \reaction^{w_{nm}}_{w_{mn}} m + \sa \s\A.
  \label{reaction}
\end{equation}
Here, $\r,\s$ are the numbers of species $\A$ involved in this transition, see
Fig.~\ref{figure2}. We assume a dilute solution of $\A$ molecules in a solvent
(modeled as a heat bath at constant temperature). Reaction time constants
should thus be much larger than diffusion time constants. Hence, mass action
law kinetics with respect to the $\A$ molecules is a good approximation and
the ratio between forward rate $\wnm$ and backward rate $\wmn$ is given by
\begin{equation}
  {\wnm\over\wmn} = {\wnmo\over \wmno} \prod_\alpha (\c)^{\r-\s}.
  \label{eq:w}
\end{equation} 
Here, we separate the concentration dependence from some ``intrinsic'' or bare
rates $\wnmo,\wmno$. Their ratio can be determined by considering a
hypothetical equilibrium condition for this reaction. In fact, if the reaction
took place in equilibrium with concentrations $\ceq$, we would have the
detailed balance relation
\begin{equation}
  {\wnm^{\rm eq}\over\wmn^{\rm eq}} = 
  {\wnmo\over \wmno} \prod_\alpha {\ceqa} = {p^{\rm eq}_m \over p^{\rm eq}_n}
  =\exp \left(-\Delta G\right)  
  \label{eq:wo},
\end{equation}
where 
\begin{equation}
  \Delta G \equiv-[ E_n - E_m +\sa \d \meq]
  \label{eq:G}
\end{equation}
is the equilibrium free energy difference for this reaction and $p^{\rm
  eq}_{m,n}$ are the equilibrium probabilities of states $m$ and $n$,
respectively. The chemical potential for species $\alpha$ quite generally
reads
\begin{equation}
  \m\equiv E_\alpha + \ln \c
\end{equation}
which for equilibrium becomes $\meq=E_\alpha + \ln \ceq$. Combining this with
Eqs.~(\ref{eq:wo}) and~(\ref{eq:G}) shows that the ratio of the intrinsic
rates
\begin{equation}
  {\wnmo\over \wmno} = \exp[E_n-E_m + \sa\d\E]
  \label{eq:E}
\end{equation}
involves only the energy-terms and is independent of concentrations.
Eq.~(\ref{eq:w}) for the ratio under non-equilibrium conditions then becomes
\begin{equation}
  \ln{\wnm\over\wmn} = E_n-E_m+\sum_{\alpha}\d\m \equiv -\Delta E+\wc^{nm}.
  \label{eq:1}
\end{equation}
The right hand side corresponds to the difference between applied chemical
work
\begin{equation}
  \wc^{nm}=\sum_{\alpha}\d\m  
\end{equation}
(since every transformed $\A$ molecule gives rise to a chemical work $\m$) and
the difference in internal energy $\Delta E$. For the first law to hold for
this transition, we then have to identify the left hand side of
Eq.~\eqref{eq:1} with the heat delivered to the medium, i.e. with the change
in entropy of the medium
\begin{equation}
  \ln {\wnm\over\wmn} = \Delta s^{nm}_{\rm m}.
  \label{eq:h}
\end{equation}

We now show that this identification between the ratio of the forward rate and
the backward rate with the heat exchanged in this step and hence the change in
entropy of the medium (arising here from an interpretation of a single
reaction step in terms of the first law) fits into the general scheme of
entropy production in stochastic dynamics introduced in Ref.~\cite{seif05a}.

\subsection{Entropy production in stochastic network dynamics}

We briefly recall the essential relations of Ref.~\cite{seif05a} where entropy
production was defined quite generally for a Markovian dynamics on a discrete
set of states $\{n\}$. Let a transition between discrete states $n$ and $m$
occur with a rate $w_{nm}(\l)$, which depends on an externally controlled
time-dependent parameter $\l(\tau)$. The master equation for the
time-dependent probability $p_n(\tau)$ then reads
\begin{equation}
  \partial_\tau p_n(\tau) = \sum_{m \not = n}
  [w_{mn}(\l) p_m(\tau) - w_{nm}(\l) p_n(\tau)].
  \label{eq:me}
\end{equation}
For any fixed $\l$, there is a steady state $\ps_n(\l)$~\cite{schn76}.

A stochastic trajectory $n(\tau)$ starts at $n_0$ and jumps at times $\tau_j$
from $n_j^-$ to $n_j^+$ ending up at $n_t$. As entropy along this trajectory,
we have defined
\begin{equation}
  s(\tau)\equiv - \ln \pnt,
  \label{eq:ent}
\end{equation}
where $\pnt$ is the solution $p_n(\tau)$ of the master equation~(\ref{eq:me})
for a given initial distribution $p_n(0)$ taken along the specific trajectory
$n(\tau)$. The rate of entropy flow into the medium is defined as
\begin{equation}
  \dsm\equiv
  \sum_j\delta(\tau-\tau_j) 
  \ln{\wmp\over\wpm}
  \equiv\sum_j\delta(\tau-\tau_j)\Delta s^{n_j^-n_j^+}_{\rm m},
  \label{eq:sm}
\end{equation}
which leads to a change in the medium entropy along a trajectory of length $t$
as
\begin{equation}
  \Delta s_{\rm m} = \IInt{\tau}{0}{t} \dsm.
  \label{eq:sm:delta}
\end{equation}
The total entropy change
\begin{equation}
  \Delta s\t \equiv s(t)-s(0) + \Delta s_{\rm m}
  = \sum_j\ln\frac{p_{n_j^-}}{p_{n_j^+}} + \sum_j \ln{\wmp\over\wpm}
  \label{eq:stot}
\end{equation}
then obeys an integral fluctuation theorem
\begin{equation}
  \mean{\exp[-\Delta s\tot]} = 1,
  \label{eq:R3}
\end{equation}
which implies the second law like statement
\begin{equation}
  \mean{\Delta s\t} \geq 0.
\end{equation}
Likewise, one has in complete analogy to the mechanically driven case discussed
above the extension
\begin{equation}
  \mean{f(n_t)\exp[-\Delta s\t]} =\mean{f(n_t)},
  \label{eq:R4}
\end{equation}
where $f(n_t)$ is any function of the final state.

These results hold for the non-equilibrium average with arbitrary initial
state, arbitrary time-dependent rates $w_{nm}(\l)$ caused, e.g., by
time-dependent concentrations $\c(\l)$, and any length $t$ of trajectories.

Even though the entropy definition for the system~(\ref{eq:ent}) and the
medium~(\ref{eq:sm}) have been given in Ref.~\cite{seif05a} purely formally
(or at most in analogy with the mechanically driven case), this definition of
the change in entropy of the medium~(\ref{eq:sm:delta}) corresponds exactly to
the one found in~(\ref{eq:h}) for our biomolecular example derived on the
basis of the kinetics together with the first law formulation along a
trajectory.  Crucial for this agreement, however, is the persistence of the
relation~(\ref{eq:E}) for the intrinsic rates in a non-equilibrium situation.
In fact, this persistence corresponds to maintaining the
correlations~(\ref{eq:corr}) in non-equilibrium in the mechanically driven
case.

\subsection{Several molecules or equivalent internal states: Role of
  ``degeneracy''}

The definitions~(\ref{eq:ent}) and~(\ref{eq:sm}) for system entropy and
entropy change of the medium are correct and consistent with the simple
assumptions for the kinetics if $n$ and $m$ label single states.  An important
modification arises if several states are lumped into one label $n$.

\begin{figure}
  \includegraphics[width = 7cm]{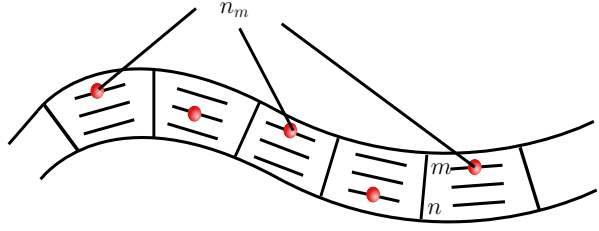}
  \caption{Sketch of a multi-domain protein with $N$ equivalent consecutive
    ``reaction sites'', each involving $M$ (here 3) internal states. The
    number of sites which are in the internal state $m$ is $n_m$ (here 3).}
  \label{figure3}
\end{figure}

As an example, consider the case of $N$ identical but spatially separable and
hence in principle distinguishable molecules of the type discussed above each
involved in reactions~(\ref{reaction}). If we can resolve only the numbers
$\n=(n_1,\dots,n_M)$ of molecules which are in a particular state but cannot
distinguish which of the $n_n$ equivalent molecules undergoes the transition
from $n$ to $m$, the state space can now be labeled by $\n$ with the
constraint $\sum_{i=1}^M n_i=N$. Likewise, we could assume we have $N$
equivalent reaction sites lined up consecutively along a multi-domain protein
where each site could be in any of the $M$ states, see Fig.~\ref{figure3}.

We now denote the rate for a transition from $\n$ to $\n'$ with
\begin{equation}
  n'_i=n_i -\delta_{in} + \delta_{im}
\end{equation}
by $\Wnm$ and the corresponding backward rate as $\Wmn$. Mass action law
kinetics implies
\begin{equation}
  {\Wnm\over\Wmn}={\wnm n_n\over\wmn(n_m+1)}
\end{equation}
since the forward rate is enhanced by the factor $n_n$ which counts the number
of molecules in the state $n$. Likewise, for the corresponding backward
transition, any of the (then) $n_m+1$ molecules in state $m$ can jump. If one
forward reaction takes place, the entropy change of the medium $\Delta s^{\bf
  n n'}_{\rm m}$ is still given by
\begin{equation}
  \Delta s^{\bf n n'}_{\rm m} = E_n - E_m + \d\m  = \ln\frac{\wnm}{\wmn},
  \label{eq:smdg}
\end{equation} 
since the first law for a single reaction event remains the same as above. On
the other hand, by naive application of the general expression~(\ref{eq:h}) as
\begin{equation}
  \Delta s^{\bf n n'}_{\rm m} = \ln\frac{\Wnm}{\Wmn}
  = \Delta s^{nm}_{\rm m} + \ln\frac{n_n}{n_m+1}
\end{equation}
one would obtain an additional term $\ln [n_n/(n_m+1)]$.

The solution of this apparent inconsistency requires an analysis of the
entropy definition~(\ref{eq:ent}) in the case of degeneracy. In our example,
the state $\n$ carries a degeneracy
\begin{equation}
  g_{\n}={N!\over\prod_i n_i!}  .
\end{equation} 
We now define the stochastic entropy of the state $\n$ not by~(\ref{eq:ent})
but rather by
\begin{equation}
  s(\tau)\equiv -\ln p_{\n(\tau)}(\tau) + s_{\n (\tau)}^0
  \label{eq:dent}
\end{equation}
with the ``intrinsic'' entropy
\begin{equation}
  s_\n^0\equiv \ln g_\n  
\end{equation}
determined by the degeneracy. For a single transition $\n$ to $\n'$ at time
$\tau$, the system entropy then changes according to
\begin{equation}
  \label{eq:sm:mod}
  \Delta s^{\bf n n'} = \ln\frac{p_{\n}(\tau)}{p_{\n'}(\tau)} 
  + \ln\frac{g_{\n'}}{g_{\n}}
  = \ln\frac{p_{\n}(\tau)}{p_{\n'}(\tau)} + \ln\frac{n_n}{n_m+1}.
\end{equation}
If we use this modified definition of system entropy change~(\ref{eq:sm:mod})
and the thermodynamically correct change in medium entropy~(\ref{eq:smdg}),
the total entropy production in a single step
\begin{equation}
  \Delta s^{\bf nn'}_{\rm tot} = \Delta s^{\bf nn'} + \Delta s^{\bf nn'}_{\rm m}
  = \ln\frac{p_{\n}(\tau)}{p_{{\bf n'}}(\tau)} + \ln \frac {\Wnm}{\Wmn}
\end{equation}
has the form of the right hand side of Eq.~(\ref{eq:stot}). Hence, the
fluctuation theorems~(\ref{eq:R3}) and~(\ref{eq:R4}) even hold in the case of
a degenerate state space.

Generalizing and summarizing this procedure, we modify the expression
developed in Ref.~\cite{seif05a} for the change of the medium entropy as
\begin{equation}
  \dsm\equiv
  \sum_j\delta(\tau-\tau_j) 
  \left[\ln{\wmpb\over\wpmb}-(s^0_{\njp}-s^0_{\njm}) \right],
  \label{eq:sm2}
\end{equation}
where the additional term in round brackets compensates for each jump the
change in the degeneracy factor. In the example discussed above, we now get
for the contribution of this transition to the change in medium entropy
\begin{equation}
  \Delta s^{\bf n n'}_{\rm m} \equiv \ln{\Wnm\over \Wmn} - 
  \ln{g_\n\over g_{\n'}} = \ln {\wnm\over \wmn},
\end{equation}
which is indeed the thermodynamically correct expression. Hence, the modified
definitions~(\ref{eq:dent}) and~(\ref{eq:sm2}) for system and medium entropy
change in the presence of degeneracy are not only consistent with a first
law-like energy conservation but also obey the fluctuation theorems. While we
have identified the intrinsic entropy with the degeneracy, it is tempting to
speculate that even for other sources of intrinsic entropy the
definitions~(\ref{eq:dent}) and~\eqref{eq:sm2} remain meaningful.

\subsection{Detailed fluctuation theorem in the steady state}

The reaction network discussed above allows also for a genuine non-equilibrium
steady state. Necessary for such a state are at least three internal states in
order to have at least one cycle in the network, i.e. two essentially
different reaction paths leading to the same final state. A non-equilibrium
steady state can be obtained if it is possible to adjust the concentrations
$\{c_\alpha\}$ such that a net flux in the species $A_\alpha$ occurs. Hence,
the stationary state violates the detailed balance condition $p^{\rm s}_n \wnm
= p^{\rm s}_m \wmn$. For such non-equilibrium steady states a detailed
fluctuation theorem
\begin{equation}
  p(-\Delta s_{\rm tot}) = \exp[-\Delta s_{\rm tot}] p(\Delta s_{\rm tot})
  \label{eq:detFT}
\end{equation}
holds with the present entropy definition for any length of the
trajectory~\cite{seif05a} thus extending previous results valid in the
long-time limit~\cite{evan94,gall95,kurc98,lebo99,maes03}.


\section{Summarizing perspectives}
\label{sec:persp}

The thermodynamically consistent description of non-equilibrium processes of
small systems developed in this paper paradigmatically relies on two central
concepts. First, we need a first-law like energy balance along the stochastic
trajectory. While its form is pretty obvious in the mechanical case, it is
less straightforward in the chemical case where it involves identifying the
dissipated heat as the ratio of the forward and backward rate (up to a
possible degeneracy correction). Second, the non-equilibrium dynamics has to
be formulated in such a way that if it is restricted to the equilibrium
concentrations it obeys detailed balance with the appropriate equilibrium
distribution. This condition does not determine the non-equilibrium dynamics
uniquely. Still, the present choice for the rates both in the mechanical and
in the chemical case seems to be the ``minimal'' extension of the equilibrium
rates. In fact, one could call such a dynamics an {\em isothermal
  non-equilibrium dynamics} since the notion of temperature of the surrounding
heat bath still makes sense and serves to relate exchanged heat (occurring in
the first law) with an entropy change of the medium (entering the second law).
For this type of dynamics, entropy along a stochastic trajectory can
consistently be defined such that (i) it reduces upon averaging to the usual
non-equilibrium ensemble formulation; and (ii) together with the
identification of the entropy change of the medium the total entropy change
obeys exact relations from which a second law for the average follows
trivially.

Combining the chemically driven with the mechanically driven case discussed
here separately is straightforward. Along this line, one could then apply our
concepts to models previously introduced to describe such coupled systems like
in Refs.~\cite{qian02,brau05} or the motor models mentioned in the
introduction. Likewise, the chemically driven case discussed here for one (or
several identical) reaction sites can be extended to a thermodynamically
consistent theory of any small-scale biochemical reaction network as will be
discussed elsewhere~\cite{schm06}.

The theoretical framework developed in this paper is quite general. Leaving
the appeal of exact relations aside, its significance for any specific system
will depend on working out the particular details. Of special interest seem to
be the distribution for the entropy changes of system, medium and their sum.
Presumably only little can be said for these distribution in general since
even for simple driven non-biological two-level systems these distributions
can exhibit a quite rich structure~\cite{schu05}. For a simple three-state
model of the rotary motor in the steady state, the exact distribution of the
entropy change is available through mapping to an asymmetric random
walk~\cite{seif05}. Numerical analysis of more sophisticated models should
finally provide us with a better understanding of how entropy changes on the
stochastic level look like beyond the exact constraints developed in this
paper. Finally, it will be exciting to see when and how these elements of a
non-equilibrium thermodynamics will be integrated to a consistent and
comprehensive theory of the physics of the cell.


\appendix

\section{Proof of integral fluctuation theorems}
\label{sec:ft}

In this appendix we show how to extend proofs~\cite{croo00,maes03,seif05a} of
integral fluctuation relations based on time-reversal to many degrees of
freedom involving hydrodynamic interactions. The integral fluctuation theorem
for the total entropy production~\eqref{eq:ft}, the Jarzynski relation, and
the more general relation~\eqref{eq:ft:cro} then all derive from one master
formula, which has been given before for the one-dimensional case in
Ref.~\cite{seif05a}.

Since the thermal noise $\zeta_i(\tau)$ in Eq.~\eqref{eq:lang} is modeled as
Gaussian noise, the probability for a noise trajectory is
$P[\nt(\tau)]=\NN\exp\{-A[\nt(\tau)]\}$ with ``action''
\begin{equation}
  \label{eq:apx:action}
  A[\nt(\tau)] \equiv \frac{1}{2}\IInt{\tau}{0}{t}\IInt{\tau'}{0}{t}
  \zeta_i(\tau)K^{-1}_{ij}(\tau-\tau')\zeta_j(\tau'),
\end{equation}
correlation matrix
$K_{ij}(\tau-\tau')\equiv\mean{\zeta_i(\tau)\zeta_j(\tau')}$, and
normalization $\NN$. We make the transition from the noise history $\nt(\tau)$
to the trajectory $\x(\tau)$ given the initial state $\x_0$ by inserting the
Langevin equation~\eqref{eq:lang}
\begin{equation}
  {\dot x}_i = -\mu_{ij}(\x)\pd{V}{x_j}(\x,\l(\tau)) + \zeta_i
  \equiv v_i(\x,\tau)+\zeta_i
\end{equation}
along with the noise correlation~\eqref{eq:corr} into
Eq.~\eqref{eq:apx:action}, leading to
\begin{equation}
  \label{eq:apx:act}
  A[\nt(\tau)] = \frac{1}{4}\IInt{\tau}{0}{t}
  \left[{\dot x}_i(\tau)-v_i(\x(\tau),\tau)\right]
  \mu^{-1}_{ij}\left[{\dot x}_j(\tau)-v_j(\x(\tau),\tau)\right].
\end{equation}

The change of variables from $\nt(\tau)$ to $\x(\tau)$ also leads to a
Jacobian $J[\x(\tau)]$ in the trajectory weight. The Langevin equation
discretized into $N$ steps takes the form
\begin{equation}
  \frac{x^\alpha_i-x^{\alpha-1}_i}{\eps} = \frac{1}{2}\left[
    v_i^\alpha(\xal) + v_i^{\alpha-1}(\x^{\alpha-1}) \right] + \zeta^\alpha_i,
\end{equation}
where the upper Greek indices represent discrete time and $\eps$ is a small
time step. This discretization corresponds to Stratonovich's scheme. The
Jacobian matrix resulting from the change of variables is
\begin{equation}
  J^{\alpha\beta}_{ij} \equiv \pd{\zeta^\alpha_i}{x^\beta_j},
\end{equation}
from which we calculate the Jacobian as
\begin{equation}
  J[\x(\tau)] \equiv \lim_{\eps\rightarrow 0}\det J^{\alpha\beta}_{ij}.
\end{equation}
In order to see the structure of the Jacobian matrix, we define for a given
time index $\alpha$
\begin{equation}
  \pm M^\alpha_{ij} \equiv \pm\delta_{ij} 
  -\frac{\eps}{2}\pd{v^\alpha_i}{x_j}(\x^\alpha)
  \approx\pm
  \left[\exp\left\{\mp\frac{\eps}{2}\pd{v^\alpha_k}{x_l}(\x^\alpha)\right\}
  \right]_{ij}.
\end{equation}
The Jacobian matrix can then be written as matrix of matrices
\begin{equation}
  \J = \frac{1}{\eps}\left\lgroup
    \begin{array}{ccccl}
       +\M^1 &     0 &     0 &     0 & \\
       -\M^1 & +\M^2 &     0 &     0 & \\
           0 & -\M^2 & +\M^3 &     0 & \\
           0 &     0 & -\M^3 & +\M^4 & \\
             &       &       &       &\ddots \\
    \end{array}
  \right\rgroup_{N\times N},
\end{equation}
from which the determinant immediately follows as
\begin{equation}
  J[\x(\tau)] = \lim_{\eps\rightarrow 0}\eps^{-Nd}\prod_{\alpha=1}^N \det\M^\alpha.
\end{equation}
Using the identity $\det\exp=\exp\tr$ and after taking the limit
$\eps\rightarrow0$, $N\rightarrow\infty$ with $N\eps=t$ we finally arrive at
\begin{equation}
  J[\x(\tau)] = \exp\left(
    -\frac{1}{2}\IInt{\tau}{0}{t}\sum_{ij}\pd{v_i}{x_j}(\x(\tau),\tau)\right).
\end{equation}

The action~\eqref{eq:apx:act} along a stochastic trajectory can be split into
two contributions
\begin{eqnarray}
  \label{eq:apx:action:s}
  A_\mathrm{s}[\x(\tau)|\x_0] &=& \frac{1}{4}\IInt{\tau}{0}{t} \left\{
    \dot x_i\mu^{-1}_{ij}\dot x_j + \pd{V}{x_i}\mu_{ij}\pd{V}{x_j} \right\}, \\
  \label{eq:apx:action:a}
  A_\mathrm{a}[\x(\tau)|\x_0] &=& \frac{1}{2}\IInt{\tau}{0}{t}
  \pd{V}{x_i}\dot x_i = -\frac{\Delta s_\mathrm{m}}{2},
\end{eqnarray}
$A=A_\mathrm{s}+A_\mathrm{a}$, where for the last equality we have used
Eq.~\eqref{eq:sm:mech}. Under time reversal, i.e., under the transformation
\begin{equation}
  \tau \mapsto t-\tau \equiv \tilde\tau:\ 
  \lambda(\tau) \mapsto \tilde\lambda(\tilde\tau), \quad
  x_i(\tau) \mapsto \tilde x_i(\tilde\tau), \quad
  \dot x_i(\tau) \mapsto -\dot{\tilde x}_i(\tilde\tau)
  \label{eq:apx:trafo}
\end{equation}
the symmetric part of the action stays invariant, $\tilde
A_\mathrm{s}=A_\mathrm{s}$, whereas $\tilde A_\mathrm{a}=-A_\mathrm{a}$
changes sign. Since the Jacobian $J$ only involves mobility $\mu$ and
potential energy $V$ it is invariant under time reversal, $\tilde J=J$.  For
given initial state $\x_0$ and final state $\x_t=\tilde\x_0$, the total
trajectory weight becomes
\begin{eqnarray}
  P[\x(\tau)|\x_0] &=&
  \NN J[\x(\tau)|\x_0]\exp\left\{-A_\mathrm{s}[\x(\tau)|\x_0]-
  A_\mathrm{a}[\x(\tau)|\x_0]\right\},\\
  \tilde P[\tilde\x(\tilde\tau)|\tilde\x_0]
  &=& \NN J[\x(\tau)|\x_0]\exp\left\{-A_\mathrm{s}[\x(\tau)|\x_0]+
  A_\mathrm{a}[\x(\tau)|\x_0]\right\}.
\end{eqnarray}

In order to prove a general version of the integral fluctuation theorem we
combine the physical picture of time reversal with a generalization of the
actual final distribution $p(\x_t)$ to an arbitrary normalized initial
distribution $p_1(\tilde\x_0)$ for time-reversed paths. Normalization then
implies
\begin{equation}
  1 = \sum_{\tilde\x(\tau)}\tilde
  P[\tilde\x(\tilde\tau)|\tilde\x_0]p_1(\tilde\x_0),
\end{equation}
where the summation runs over all trajectories. Inserting the actual initial
distribution $p_0(\x_0)$ we have the master formula
\begin{equation}
  1 = \sum_{\x(\tau)}\frac{
    \tilde P[\tilde\x(\tilde\tau)|\tilde\x_0]p_1(\tilde\x_0)}{
    P[\x(\tau)|\x_0]p_0(\x_0)}P[\x(\tau)|\x_0]p_0(\x_0)
  = \mean{\frac{
      \tilde P[\tilde\x(\tilde\tau)|\tilde\x_0]p_1(\tilde\x_0)}{
      P[\x(\tau)|\x_0]p_0(\x_0)}}
  \equiv \mean{\exp[-R]}
\end{equation}
with
\begin{equation}
  R = \ln\frac{P[\x(\tau)|\x_0]p_0(\x_0)}
  {\tilde P[\tilde\x(\tilde\tau)|\tilde\x_0]p_1(\tilde\x_0)}
  = -\ln\frac{p_1}{p_0}+\Delta s_\mathrm{m}.
\end{equation}
Replacement of $\sum_{\tilde\x}$ by $\sum_{\x}$ is admissible since it does
not matter how we denote the summation variable when we sum over all
trajectories.

For the proof of Eq.~\eqref{eq:ft}, we choose with $p_1(\x)=p(\x,t)$ the
actual probability distribution at the end of the trajectory. With $p_0(\x)$
the distribution of the initial state, the ratio
\begin{equation}
  R = \Delta s + \Delta s_\mathrm{m} = \Delta s\t
\end{equation}
becomes the total change of entropy. If we choose the normalized function
\begin{equation}
  p_1(\x) = \frac{f(\x)p(\x,t)}{\mean{f(\x)}}
\end{equation}
with an arbitrary function $f(\x)$, where the average in the denominator
implies the distribution $p(\x,t)$, the ratio becomes
\begin{equation}
  R = \Delta s\t - \ln\frac{f(\x)}{\mean{f(\x)}},
\end{equation}
leading to Eq.~\eqref{eq:ft:cro}. Finally, if one chooses $p_1(\x)=p_{\rm
  eq}(\x,\l(t))$, one obtains the Jarzynski relation~\eqref{eq:ft:jarz} and
analogously the general relation~\eqref{eq:ft:crooks}.


\end{document}